\documentclass{phbauth}        
\usepackage{graphicx}

\newcommand{\UTB}{U$_{.9725}$Th$_{.0275}$Be$_{13}$}

\begin{document}

\begin{frontmatter}

\title{Unconventional strong pinning in the low 
temperature phase of U$_{.9725}$Th$_{.0275}$Be$_{13}$}

\author[address1]{Elisabeth Dumont\thanksref{thank1}},
\author[address1]{Ana Celia Mota}, and
\author[address2]{James L. Smith}

\address[address1]{Laboratorium f\"ur Festk\"orperpkysik, ETH H\"onggerberg, 
8093 Z\"urich, Switzerland}
\address[address2]{Los Alamos National Laboratory, Los Alamos, New Mexico 
87545, USA}

\thanks[thank1]{Corresponding author. Present address: 
Laboratorium f\"ur Festk\"orperpkysik, ETH H\"onggerberg, CH 8093 
Z\"urich, Switzerland. E-mail: dumont@solid.phys.ethz.ch} 

\begin{abstract}

We investigated low field vortex dynamics in a single crystal of 
U$_{.9725}$Th$_{.0275}$Be$_{13}$.We found a sharp transition in 
the vortex creep rate at the lower transition temperature $T_{c2}$, 
coincident with the second jump in the specific heat. In the 
high-temperature phase, rather strong creep rates are observed. 
In the low temperature phase, the rates drop to undetectabely low levels. 
This behaviour indicates that a very strong pinning mechanism is present 
in the low temperature phase of U$_{.9725}$Th$_{.0275}$Be$_{13}$, which 
could be explained by the existence of domain walls, separating discreetly 
degenerate states of a superconductor, that can sustain fractional vortices 
and thus act as very strong pinning centers. 

\end{abstract}


\end{frontmatter}


If a small amount of uranium is substituted for thorium in the heavy 
fermion superconductor UBe$_{13}$, the superconducting transition 
temperature shows a non-monotonic dependence on thorium concentration 
$x$.  Moreover, in a small critical range $0.019<x<0.045$ a second 
transition occurs below the superconducting transition\cite{ott}.  The 
nature of this phase transition is still a matter of controversy.  In 
zero field $\mu$SR measurements, a continuous increase of 
$\mu$-spin-relaxation rate, which sets in at $T_{c2}$ has been 
observed\cite{muSR_UTB}.  This has been interpreted as originating 
from a very weak spontaneous magnetic field.  Superconducting origin 
of the transition at $T_{c2}$ has been deduced from the observation of 
a distinct increase in the slope of the lower critical field 
$H_{c1}$\cite{Hc1_UTB} at $T_{c2}$.

Recently, the low-field magnetic properties of the heavy fermion 
superconductor UPt$_{3}$ have been investigated \cite{andreas}.
In the low--field, low--temperature B--phase, 
bulk vortices remain so strongly pinned that the creep rates drop from 
rather high finite rates observed in the A--phase to undetectabely low 
levels at $T_{c}^{-}$.  
The observation of zero initial creep in the B--phase of UPt$_{3}$
was then interpreted as originating from an intrinsic pinning 
mechanism in multicomponent superconducting phases.  In such phases 
fractional vortices get trapped in domain walls separating discreetly 
degenerate states of the superconductor\cite{sigrist}.  These 
experimental results show that vortex creep measurements can be used 
as a powerful probe to give information about the character of a 
superconducting phase.

\begin{figure}[h]
\begin{center}\leavevmode
\includegraphics{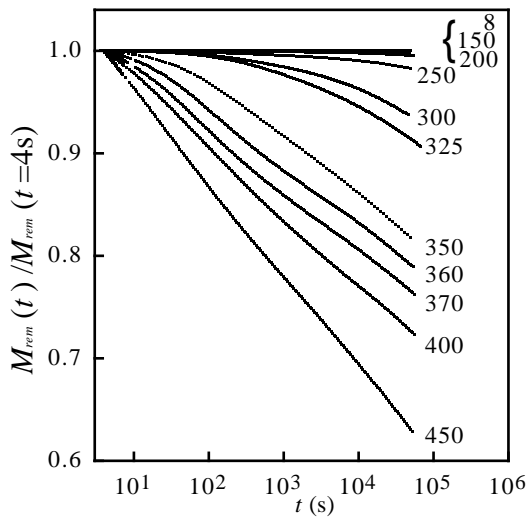}
\caption{Isothermal relaxation measurements of remanent $M_{rem}$ in \UTB. 
Temperatures are given in~mK on the right hand side of the graph. }
\label{decays}
\end{center}
\end{figure}

We present here relaxation measurements of the remanent magnetization 
on a single crystal of \UTB (size: $1.00\times 0.89 \times 2.25$\,mm$^3$) 
prepared in Los Alamos National Laboratory\cite{smith}.  The experimental 
arrangement has been described earlier \cite{andreas}.  The single 
crystal has a transition temperature $T_{c1}=523$\,mK. Specific heat 
measurements \cite{art} show a second jump at $T_{c2}=350$\,mK. 
Typical relaxation measurements of the remanent magnetization are 
shown in fig.~\ref{decays}.  All the values of $M_{rem}$ are taken 
with the sample in the fully critical state.

\begin{figure}[h]
\begin{center}\leavevmode
\includegraphics{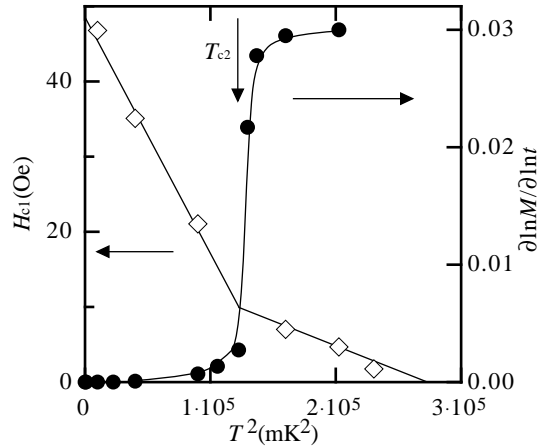}
\caption{Creep rates (closed circles, right axis) and lower critical 
field $H_{c1}$ (open diamonds, left axis) of \UTB plotted vs $T^2$.  
The lines are guides to the eye.}
\label{rates}
\end{center}
\end{figure}

We observe two different regimes of vortex creep, separated by a 
dramatic change in creep rates at $T=T_{c2}$.  For temperatures above 
$T_{c2}$, we observe strong vortex creep and a logarithmic time 
dependence of the decays.  For temperatures below $T_{c2}$, the 
initial vortex creep drops to zero.  At longer times the decays start 
deviating from logarithmic behaviour.  This acceleration of vortex 
creep at long times has also been observed in UPt$_{3}$\cite{andreas}.  
Here we only discuss the initial slope of the decays.  The normalized 
creep rates ($\partial \ln M/ \partial \ln t$) are plotted in 
Fig.\ref{rates} as a function of temperature.  At $T=T_{c2}$, we 
observe a sharp transition in creep rates.  They drop by 3 orders of 
magnitude to zero within our sensitivity ($\partial \ln M/ \partial 
\ln t \approx 10^{-5}$).  Also shown in Fig.\ref{rates} are 
measurements of the lower critical field $H_{c1}$ on the same crystal.  
$H_{c1}$ was obtained as the field where the first deviation from the 
initial slope of isothermal magnetization--curves occurs.

In conclusion, we have observed a sharp transition in vortex pinning 
in \UTB at $T_{c2}$, similar to the one observed in UPt$_{3}$ at 
$T^{-}_{c}$\cite{andreas}.  This might hint towards a new type of 
strong pinning  structures and/or a new type of vortices being present 
in the low--T phases of these two multiphase superconductors.


\end{document}